%% file: cikm2016-evaluation.tex
\documentclass{sig-alternate}

\CopyrightYear{2016}
\setcopyright{rightsretained}
\conferenceinfo{CIKM'16 }{October 24-28, 2016, Indianapolis, IN, USA}
\isbn{978-1-4503-4073-1/16/10}
\doi{http://dx.doi.org/10.1145/2983323.2983829}

\usepackage{algorithm}
\usepackage[noend]{algpseudocode}
\usepackage{amssymb}
\usepackage{amsmath}
\usepackage{booktabs}
\usepackage{caption}
\DeclareCaptionType{copyrightbox}
\usepackage{color}
\usepackage{dashrule}
\usepackage{paralist}
\usepackage{enumitem}
\setlist{nosep}
\usepackage{graphicx}
\usepackage{mathabx}
\usepackage{mdwlist}
\usepackage[square,comma,numbers,sort&compress,sectionbib]{natbib}
\usepackage[group-separator={,},group-minimum-digits=4]{siunitx}
\usepackage{subfig}
\usepackage{tabularx}
\usepackage{textcomp}
\usepackage{tikz}
\usepackage[normalem]{ulem}
\usepackage{xcolor}

\usepackage{hyperref}

\newcommand{\squishlist}{
 \begin{list}{$\bullet$}
  { \setlength{\itemsep}{0pt}
     \setlength{\parsep}{3pt}
     \setlength{\topsep}{3pt}
     \setlength{\partopsep}{0pt}
     \setlength{\leftmargin}{1.5em}
     \setlength{\labelwidth}{1em}
     \setlength{\labelsep}{0.5em} } }

\newcommand{\squishlisttwo}{
 \begin{list}{$\bullet$}
  { \setlength{\itemsep}{0pt}
     \setlength{\parsep}{0pt}
    \setlength{\topsep}{0pt}
    \setlength{\partopsep}{0pt}
    \setlength{\leftmargin}{2em}
    \setlength{\labelwidth}{1.5em}
    \setlength{\labelsep}{0.5em} } }

\newcommand{\squishend}{
  \end{list}  }

\begin{document}
\title{Incorporating Clicks, Attention and Satisfaction into a Search Engine Result Page Evaluation Model}
\numberofauthors{2} \author{
% 1st. author
\alignauthor
Aleksandr Chuklin\\
    \affaddr{Google Research Europe \& University of Amsterdam}\\
    \affaddr{Z\"{u}rich, Switzerland}\\
    \email{chuklin@google.com}
% 2nd. author
\alignauthor
Maarten de Rijke\\
    \affaddr{University of Amsterdam}\\
    \affaddr{Amsterdam, The Netherlands}\\
    \email{derijke@uva.nl}
}
\maketitle
\begin{abstract}
    Modern search engine result pages often provide immediate value to users
    and organize information in such a way that it is easy to navigate.
    The core ranking function contributes to this and so do result snippets,
    smart organization of result blocks and extensive use of one-box answers or side panels.
    While they are useful to the user and help search engines to stand out,
    such features present two big challenges for evaluation.
    First, the presence of such elements on a search engine
    result page (SERP) may lead to the absence of clicks, which is, however,
    not related to dissatisfaction, so-called ``good abandonments.''
    Second, the non-linear layout and visual difference of SERP items
    may lead to non-trivial patterns of user attention,
    which is not captured by existing evaluation metrics.

    In this paper we propose a model of user behavior on a SERP that jointly captures
    click behavior, user attention and satisfaction, the CAS model,
    and demonstrate that it gives more accurate predictions of user actions
    and self-reported satisfaction than existing models based on clicks alone.
    We use the CAS model to build a novel evaluation metric that
    can be applied to non-linear SERP layouts and that can account for the utility
    that users obtain directly on a SERP\@.
    We demonstrate that this metric shows better agreement with
    user-reported satisfaction than conventional evaluation metrics.
\end{abstract}

%\category{H.3.3}{Information Storage and Retrieval}{Information Search and Retrieval}

\keywords{Evaluation; User behavior; Click models; Mouse movement; Good abandonment}

\input{cikm2016-evaluation-intro}
\input{cikm2016-evaluation-related}
\input{cikm2016-evaluation-model}
\input{cikm2016-evaluation-metric}
\input{cikm2016-evaluation-setup}
\input{cikm2016-evaluation-results}
\input{cikm2016-evaluation-discussion}
\input{cikm2016-evaluation-conclusion}

\medskip\noindent%
{\small%
{\bf Acknowledgements.}
This research was supported by
Ahold,
Amsterdam Data Science,
Blendle,
the Bloomberg Research Grant program,
the Dutch national program COMMIT,
Elsevier,
the European Community's Seventh Framework Programme (FP7/\-2007-2013) under
grant agreement nr 312827 (VOX-Pol),
the ESF Research Network Program ELIAS,
the Royal Dutch Academy of Sciences (KNAW) under the Elite Network Shifts project,
the Microsoft Research Ph.D.\ program,
the Netherlands eScience Center under project number 027.\-012.\-105,
the Netherlands Institute for Sound and Vision,
the Netherlands Organisation for Scientific Research (NWO)
under pro\-ject nrs
727.\-011.\-005, % SEED
612.\-001.\-116, % ImFIRE
HOR-11-10, % HORIZON
640.\-006.\-013, %DADAISM
612.\-066.\-930, %Floor
CI-14-25, %MediaNow
SH-322-15, %Cartesius
652.\-002.\-001, % Re-Search
612.\-001.\-551, % Contextual Learning to Rank for Information Retrieval
the Yahoo Faculty Research and Engagement Program,
and
Yandex.
All content represents the opinion of the authors,
which is not necessarily shared or endorsed by
their respective employers and/or sponsors.
The choice of baseline models and the CAS model suggested here
are based on prior work and the authors' own ideas
and not on practices used by commercial companies
that the authors are affiliated with or receiving funding from.
}

\setlength{\bibsep}{0pt}
\bibliographystyle{abbrvnatnourl}
\renewcommand{\bibsection}{\section*{REFERENCES}}
{\raggedright
\bibliography{cikm2016-evaluation}
}

\appendix
\input{cikm2016-evaluation-appendix}

\end{document}

%% file: cikm2016-evaluation-intro.tex
% !TEX root = cikm2016-evaluation.tex

\section{Introduction}
\label{sec:introduction}

When looking at the spectrum of queries submitted to a web search engine, we
see a heavy head of high-frequent queries (``head queries'')
as well as a long tail of low-frequent queries (``tail queries'')~\cite{Silverstein1998}.
While a small number of head queries represent a big
part of a search engine's traffic, all modern search engines can answer
these queries quite well. In contrast, tail queries are more challenging,
and improving the quality of results returned for tail queries may help a search engine to distinguish itself from its competitors.
These queries often have an underlying \emph{informational} user need:
it is not the user's goal
to navigate to a particular website, but rather to find out some information or
check a fact. Since the user is looking for information, they
may well be satisfied by the answer if it is presented directly on a SERP,
be it inside an information panel or just as part of a good result snippet.
In fact, as has been shown by \citet{Stamou2010a}, a big portion of abandoned searches
is due to pre-determined behavior: users come to a search engine
with a prior intention to find an answer on a SERP\@.
This is especially true when considering mobile search
where the network connection may be slow or the user interface may be less convenient to use.

\begin{figure}
    \begin{center}
        {%
        \setlength{\fboxsep}{0pt}%
        \setlength{\fboxrule}{0.3pt}%
        \fbox{\includegraphics[width=\linewidth]{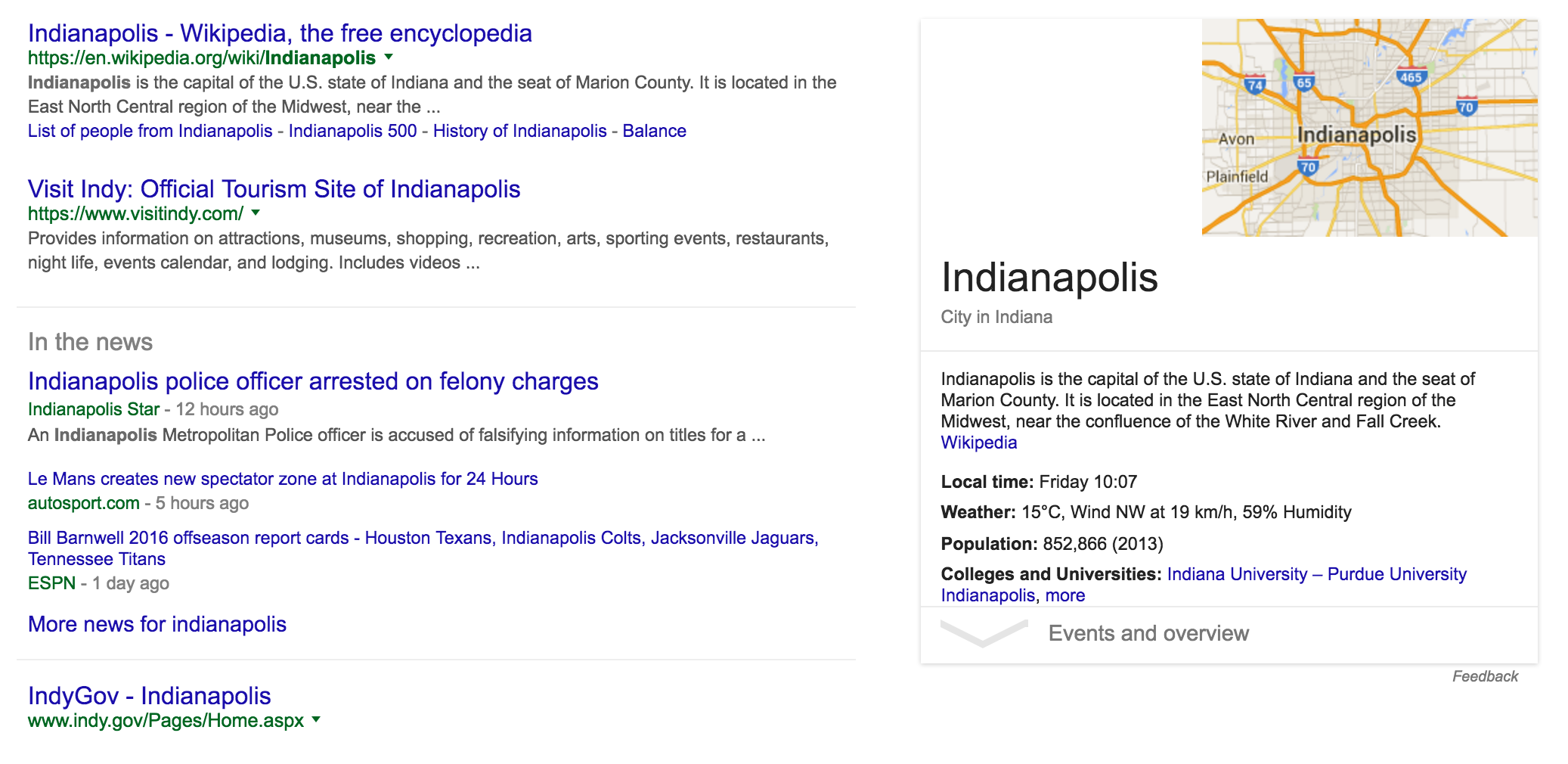}}%
        }
    \end{center}
    \caption{Example of a modern SERP with a news block and a side panel produced
    by one of the big commercial search engines for query ``Indianapolis.''}
    \label{fig:modern_serp}
\end{figure}

An important challenge arising from modern SERP layouts is that
their elements are visually different and not necessarily placed in a single column.
As was shown by~\citet{Dumais2001}, grouping similar documents helps user
to navigate faster.
Since then this approach has been studied extensively by the IR community
and adopted by the major search engines with so-called vertical blocks
and side panels (Figure~\ref{fig:modern_serp}).
When information is presented in such a way, the user examines it in a complex way,
not by simply scanning it from top to bottom~\cite{Diaz2013,Wang2013,wang-incorporating-2015}.

We claim that the currently used user models and corresponding evaluation metrics
have several disadvantages. First, most of the models assume that the SERP consists of equally shaped
result blocks, often homogeneous, presented in one column, which often prevents us from accurately measuring
user attention. Second, none of the current Cranfield-style evaluation metrics
account for the fact that the user may gain utility directly from the SERP\@.
And finally, and, perhaps, the most important of all, is that the offline evaluation metrics,
although sometimes based on a user model, do not learn from the user-reported satisfaction,
but rather use ad-hoc notions for utility and effort.

In this paper we propose an offline evaluation metric that accounts for non-trivial attention
patterns of modern SERPs and the fact that a user
can gain utility not only by clicking documents, but also by simply viewing SERP items.
Our approach consists of two steps, each having value on its own:
\begin{inparaenum}[(1)]
    \item we build a unified model of a user's clicks, attention and satisfaction, the Clicks, Attention and Satisfaction (CAS) model; and
    \item we use this model to build a Cranfield-style evaluation metric (which we call the CAS metric).
\end{inparaenum}

Consequently, our research questions can be formulated as follows:
\begin{enumerate}[label=\textbf{RQ\arabic*},labelsep=*]
    \item Does a model that unites attention and click signals give more precise
        estimations of user behavior on a SERP and self-reported satisfaction?
        How well does the model predict click vs.~satisfaction events? \label{rq1}
    \item Does an offline evaluation metric based on such a model show higher agreement with
        user-reported satisfaction than conventional metrics such as DCG\@?
        %How can this metric be extended to new unseen SERP layouts?
        \label{rq2}
\end{enumerate}

\noindent%
The rest of the paper is organized as follows.
In Section~\ref{sec:related} we discuss related work.
Then we present our user model in Section~\ref{sec:model}.
In Section~\ref{sec:metric} we present an evaluation metric based on this model.
Section~\ref{sec:experiment} describes our experimental setup.
In Section~\ref{sec:results} we present results of our experiments
followed by a discussion in Section~\ref{sec:discussion}.
We conclude in Section~\ref{sec:conclusion}.

% section introduction (end)

%% file: cikm2016-evaluation-related.tex
% !TEX root = cikm2016-evaluation.tex

\section{Related Work}
\label{sec:related}

\subsection{Abandonment}
\citet{Turpin2009} show that perceived relevance of the search results
as seen on a SERP (snippet relevance or direct SERP item relevance as we call it)
can be different from the actual relevance
and should affect the way we compute utility of the page.
\citet{Li2009} introduce the notion of \textit{good abandonment} showing
that utility can be gained directly from the SERP without clicks.
\citet{Chuklin2012b} demonstrate that the presence of snippets answering
the user query increases the number of abandonments, suggesting that the user can be satisfied
without a click. A similar study has been carried out for mobile search by~\citet{Arkhipova2014},
who perform online experiments and demonstrate that satisfaction may come
from snippets, not just from clicked results.

\subsection{Mouse movement}
Another important part of related studies concerns \textit{mouse movement}.
It has been demonstrated that there is a strong relation between mouse movement
and eye fixation, although this relation is not trivial~\cite{Rodden2008}.
Even though the correlation between eye fixation and mouse movement is far from perfect,
the latter has been shown to be a good indicator of user attention~\cite{Navalpakkam2012},
comparable in quality to eye gaze data.
In later work \citet{Navalpakkam2013} show that mouse movements are not always aligned
with eye fixations, suggesting the idea that this behavior is
user-dependent.
Based on the idea of eye-mouse association, a classifier has been developed that
can predict the fact of an individual user carefully reading a SERP item~\cite{Liu2014}
and even the satisfaction reported by the user~\cite{Liu2015}, based on mouse movements.
\citet{Huang2011} demonstrate that mouse movements can serve as a strong signal
in identifying good abandonments.
\citet{Diriye2012} show that
mouse movement data together with other signals can indeed yield an efficient
classifier of good abandonments.
Their work also introduces an experimental setup for in-situ collection
of good abandonment judgements. They argue that this is the only way
of collecting ground truth data, as even query owners have difficulties
telling the reason for abandonment if they are asked later.
%Alternatively, \citet{Hassan2013} use
%an indirect way of collecting good abandonment labels by using search engine
%switching events and post-switch predictors of dissatisfaction.

\subsection{Click models}
Previous work on \textit{click models} is also important for our study.
A \textit{click model} is a probabilistic graphical model used to predict
user clicks and in some cases even user satisfaction~\cite{Chuklin2015}.
\citet{Chuklin2013a} suggest a way to convert any conventional click model
to a Cranfield-style evaluation metric.
\citet{Huang2012} propose an extended click model that uses mouse interactions
to slightly refine an existing click model.
\citet{Chen2015} adopt a generative approach
where relevance, clicks and mousing are written as noisy functions of previous
user actions. They use this approach to predict clickthrough rates (CTRs)
of results on the SERP\@.
\citet{Diaz2013} show that visually salient SERP elements can
dramatically change mouse movement trails and suggest a model that handles this.

\medskip\noindent%
Our work is different from previous work on good abandonment in that we not only
allow for stopping after a good SERP item, but we account for this in terms of the total utility
accumulated by the user, which brings us closer to the traditional Cranfield-style evaluation approach.

Our work is different from previous work on mouse movement and click models in
that we do not study them separately, but use both as evidence for
locating the user's attention. 

On top of that, we explicitly include in our model
the notion of accumulated utility and user satisfaction as well as the possibility
to gain utility from results that were not interacted with.
% section related (end)

%% file: cikm2016-evaluation-model.tex
% !TEX root = cikm2016-evaluation.tex

\section{Model}
\label{sec:model}
Let us first describe the Clicks, Attention and Satisfaction (CAS) model that we are going to use.
It is a model of user behavior on a SERP that has three components:
\begin{itemize}
    \item an attention model;
    \item a click model; and
    \item a satisfaction model.
\end{itemize}

\begin{figure}
\centering
    \includegraphics[width=.9\columnwidth]{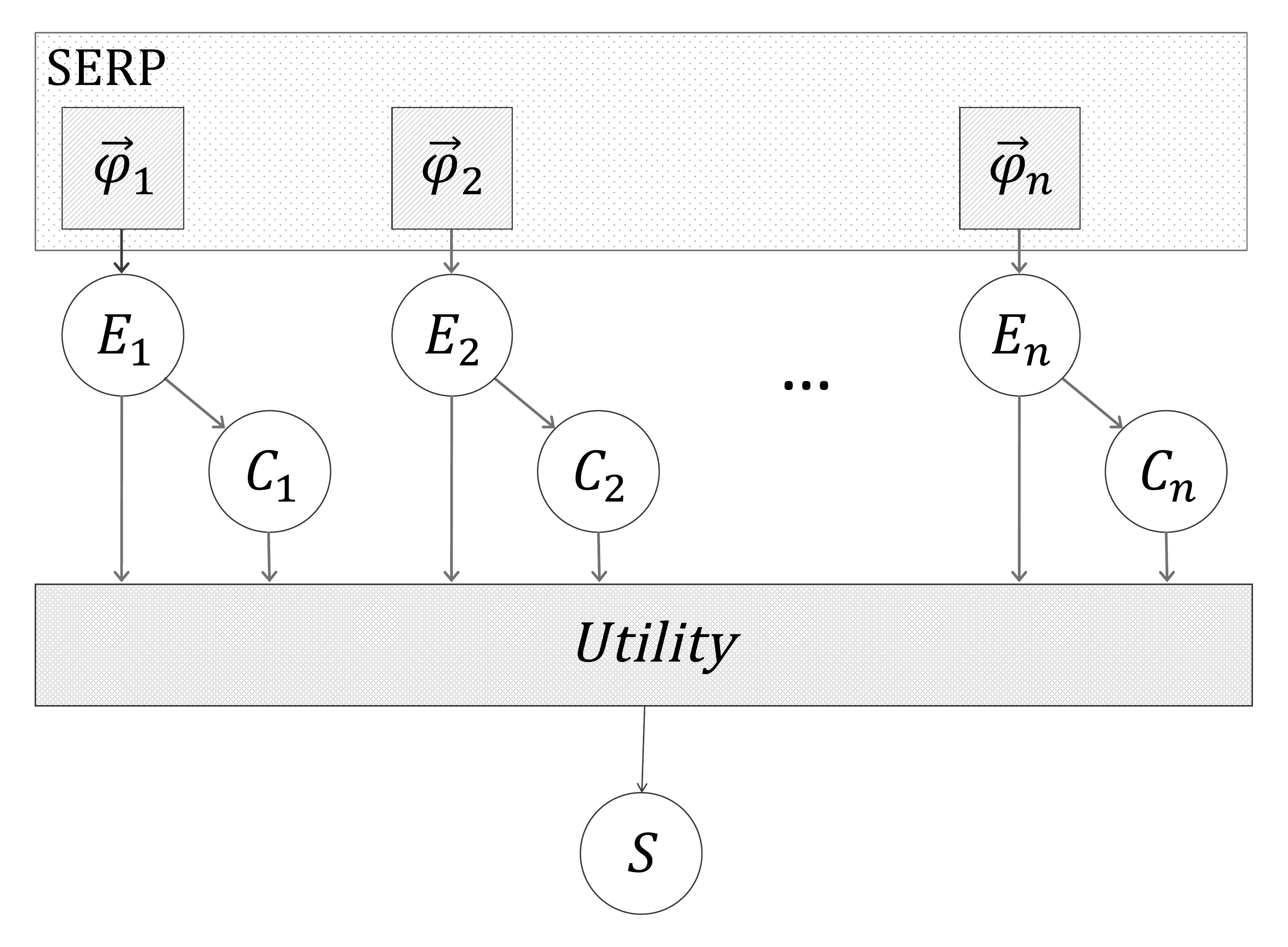}
    \vspace*{3mm}
\caption{Diagram of the CAS model.}
\label{fig:cas}
\end{figure}

\noindent%
The model is visualized in Figure~\ref{fig:cas}. Each SERP item $k$ gives rise to a feature vector $\vec{\varphi}_k$ that
determines the examination event $E_k$. After examination the user may or may not click through ($C_k$).
Then the examined and clicked documents contribute to the total utility, which, in turn determines satisfaction ($S$).
We describe each of the three components in the following sections.

We should note here that we train a \emph{relevance-based} click model, where the click probability
depends on the relevance label assigned by the raters and not on the document itself (see~\cite{Chuklin2013a}).
A classical click model can also be trained (from a bigger dataset) and compared using
click likelihood similar to what we do in Section~\ref{sub:resultsmodel}.
However, we still need a relevance-based model to build an evaluation metric (Sections~\ref{sec:metric}~and~\ref{sub:resultsmetric}).

\subsection{Attention (examination) model}
\label{sub:attentionmodel}
\citet{Diaz2013} suggest a model that predicts mouse transitions between different
elements of the SERP\@. While mousing can be used as a proxy for user attention
focus~\cite{Rodden2008,Guo2010,Navalpakkam2013}, we observe in our data entire classes of sessions where
mouse tracks and attention areas are substantially different,
while others are not.\footnote{%
For instance, currency conversion queries often result in no mousing at all,
yet the user reports satisfaction.
Similar patterns of discrepancy between mousing and attention have also been reported by~\citet{Rodden2008}.}
Hence, we cannot fully reconstruct the attention transition path. That is why, unlike~\cite{Diaz2013},
we train a \emph{pointwise} model of user attention:
\begin{equation}
    P(E_k = 1) = \varepsilon(\vec{\varphi}_k), \label{eq:exam}
\end{equation}
where $k$ is an index referring to one of the items comprising the SERP (result snippets,
weather results, knowledge panels, etc.), $E_k$ is a random variable corresponding to the
user examining item $k$, $\vec{\varphi}_k$ is a vector of features indexed by the item $k$,
and $\varepsilon$ is a function converting a feature vector into a probability. The features
we use are presented in~Table~\ref{tab:attentionfeatures}.

\begin{table}[h]
    \centering
    \caption{Features used by the attention model of CAS\@.}
    \begin{tabular}{l@{~~}p{4cm}@{~}c}
        \toprule
        Feature group & Features & \# of features \\
        \midrule
        rank & user-perceived rank of the SERP item (can be different from $k$) & 1 \\
        CSS classes & SERP item type (Web, News, Weather, Currency, Knowledge Panel, etc) & 10 \\
        geometry & offset from the top, first or second column (binary), width ($w$), height ($h$), $w \times h$ & 5 \\
        \bottomrule
    \end{tabular}
    \label{tab:attentionfeatures}
\end{table}

\noindent%
The function that converts feature vectors into probabilities is a logistic regression.
Instead of training it directly from mouse movement data, which is only a part of the examined items,
we train it in such a way that it optimizes the full likelihood of the data,
which includes not just mouse movement, but also clicks and satisfaction labels.
More on this in the following sections.
% subsection attentionmodel (end)

\subsection{Click model}
\label{sub:click_model}
For our click model we use a generalization of the Position-Based Model (PBM)~\cite{Chuklin2015},
at the core of which lies an examination hypothesis, stating that in order
to be clicked a document has to be examined and attractive:
\begin{align}
    P(C_k = 1 \mid E_k = 0) &= 0 \\
    P(C_k = 1 \mid E_k = 1) &= \alpha_{u_k}, \label{eq:attr}
\end{align}
where $C_k$ is a random variable corresponding to clicking the $k$-th SERP item,
$\alpha_{u_k}$ is the attractiveness probability of the SERP item $u_k$.
Unlike the classic PBM model, where examination is determined by the rank of the SERP item,
in our model we use a more general approach
to compute the examination probability $P(E_k = 1)$, as described in~Section~\ref{sub:attentionmodel}.
% subsection click_model (end)

\subsection{Satisfaction model}
\label{sub:satisfaction_model}
Next, we propose a satisfaction model. As we noted in the introduction,
user satisfaction may come from clicking a relevant result, but also from examining a good
SERP item. We also assume that satisfaction is not a binary event
that happens during the query session, but has a cumulative nature.
In particular, we allow the situations where after examining a good document
or a good SERP item the user may still continue the session.
This assumption is supported by data that we collected from raters (Section~\ref{sub:cfdata}).

After looking at a SERP item (referred to as ``summary extracted from a bigger document''
in the instructions), our raters were asked whether they think that
``examining the full document will be useful to answer the question $Q$''
and if so, what the reason is. While looking at the reasons specified by the raters
we found out that $42\%$ of the raters who said that they would
click through on a SERP, indicated that their goal was
``to \emph{confirm} information already present in the summary,''
which implies that the summary has an answer, yet the users continue examining it.

To put these ideas into a model, we assume that each relevant document or SERP item
that received a user's attention contributes towards the total utility $U$ gained by the user:
\begin{equation}
    U = \sum_k P(E_k = 1) u_d(\vec{D}_k) + \sum_k P(C_k = 1) u_r(\vec{R}_k), \label{eq:utility_generic}
\end{equation}
where $\vec{D}_k$ and $\vec{R}_k$ are vectors of rater-assigned labels of
direct SERP item relevance and full document relevance, respectively;
$u_d$ and $u_r$ are the transformation functions that convert the corresponding raters' labels into utility values.
To accommodate variable ratings from different raters, we assume $u_d$ and $u_r$ to be
linear functions of the rating histogram with weights learned from the data:
\begin{align}
    u_d(\vec{D}_k) &= \vec{\tau}_d \cdot \vec{D}_k \label{eq:u_d} \\
    u_r(\vec{R}_k) &= \vec{\tau}_r \cdot \vec{R}_k \label{eq:u_r},
\end{align}
where $\vec{D}_k$ and $\vec{R}_k$ are assumed to be histograms of the ratings
assigned by the raters. We have three grades for $D$ (see Figure~\ref{fig:cf_task}, question~2)
and four relevance grades for $R$ (\textit{Irrelevant}, \textit{Marginally Relevant}, \textit{Relevant}, \textit{Perfect Match});
the vectors have corresponding dimensions.

Then, we assume that the probability of satisfaction depends on the accumulated utility via the logit function:
\begin{equation}
    P(S = 1) = \sigma(\tau_0 + U) = \frac{1}{1 + e^{-\tau_0 - U}},
\end{equation}
where $\tau_0$ is an intercept.

Finally, we can write down the satisfaction probability as follows:
\begin{eqnarray}
    \lefteqn{P(S = 1) = {}} \label{eq:sat} \\
     && \sigma\left(\tau_0 + \sum_k P(E_k = 1) u_d(\vec{D}_k) + \sum_k P(C_k = 1) u_r(\vec{R}_k) \right)
     \nonumber
\end{eqnarray}
% subsection satisfaction_model (end)

\subsection{Model training}
\label{sub:training}
To be able to train the CAS model we make a further assumption that the attractiveness probability $\alpha_{u_k}$
depends only on the relevance ratings $\vec{R}_k$ assigned by the raters:\footnote{%
We also tried using separate attractiveness labels collected from the raters,
but the data was too noisy due to subjective nature of the question.
See Section~\ref{sub:cfdata} for more details.}
\begin{equation}
    P(C_k = 1 \mid E_k = 1) = \alpha(\vec{R}_k) = \sigma\left(\alpha^0 + \vec{\alpha} \cdot \vec{R}_k\right).\label{eq:alpha_logistic}
\end{equation}
Since the function $\alpha$ has to yield a probability, we set it to be a logistic regression of the rating distribution.

Now that we have the model fully specified, we can write the likelihood of the observed mouse movement, click and satisfaction
data and optimize it using a gradient descent method. We use the L-BFGS algorithm~\citep{liu1989limited}, which is often
used for logistic regression optimization. It has also been shown to be robust to correlated features~\cite{minka2003comparison}.

One important thing to note is that while computing the satisfaction probability~\eqref{eq:sat} as part of the
likelihood expression, the values of click probabilities are always either $0$ or $1$, while
the value of the examination probability can be either $1$ if there is a mouse fixation
or it is computed using~\eqref{eq:exam} if there is no mouse fixation on the SERP item.
% subsection training (end)

% section model (end)

%% file: cikm2016-evaluation-metric.tex
% !TEX root = cikm2016-evaluation.tex

\section{Search Evaluation Metric}
\label{sec:metric}
Now that we have described a model of the user's behavior on a SERP, we
can use this model to build an evaluation metric.
Once the parameters of the model are fixed, it can easily be re-used
for any new search ranking or layout change. This is very
important when working on improving a search engine and allows for quick iterations.

Assume that we have the following judgements about the SERP items
from human raters%~\cite{Chuklin2014a}
:
\begin{enumerate}
    \item direct SERP item relevance $D_k$; and
    \item topical relevance $R_k$ of the full document (assigned after clicking and examining the full document).
\end{enumerate}

\noindent%
Assume further that we have trained the model as explained in Section~\ref{sub:training}.
Now we can simply plug in the relevance labels and the model parameters
in equation~\eqref{eq:utility_generic} to obtain the utility metric:
\begin{equation}
    U = \sum_k \varepsilon(\vec{\varphi}_k) \left(u_d(\vec{D}_k) + \alpha(\vec{R}_k) u_r(\vec{R}_k) \right). \label{eq:utility}
\end{equation}

\noindent%
Note that after the parameters have been estimated and fixed,
only the raters' judgements and layout information are used to evaluate system performance.
In this way we ensure the scalability and re-usability of the Cranfield-style offline evaluation.

% section metric (end)

%% file: cikm2016-evaluation-setup.tex
% !TEX root = cikm2016-evaluation.tex

\section{Experimental Setup}
\label{sec:experiment}
Our first research question from the introduction
%(Does a model that unites attention and clicking signals
%give more precise estimations of user behavior
%on a SERP and self-reported satisfaction?
%How well does it predict click vs.~satisfaction events?)
requires us to build a model and evaluate it on self-reported satisfaction. That prompted us to collect a log of user actions.
See Section~\ref{sub:proxydata}.
Similarly, for the second question from the introduction
%(Does an offline evaluation metric based on CAS model show higher agreement
%with user-reported satisfaction than conventional metrics such as DCG?)
we need to have judgements from independent raters and we used crowdsourcing for it.
See Section~\ref{sub:cfdata}.

Below we carefully describe each step of of our data collection so as to facilitate reproducibility.
Then we detail the baseline models and the way we evaluate the models.

%Different types of data summarized in~Table~\ref{tab:datasets}.
%\begin{table*}
    %\centering
    %\caption{Different types of data collected in-situ with search proxy and
        %offline using raters.}
    %\begin{tabular}{lcccc}
        %\toprule
        %& mouse movements & clicks &  D and R labels & satisfaction  \\
        %\midrule
        %in-situ (search proxy) & \checkmark & \checkmark &           & \checkmark \\
        %\midrule
        %offline (crowdsourcing) &           &           & \checkmark & \\
        %\bottomrule
    %\end{tabular}
    %\label{tab:datasets}
%\end{table*}

\subsection{In-situ data collection}
\label{sub:proxydata}

\begin{figure}
\centering
    {%
        \setlength{\fboxsep}{0pt}%
        \setlength{\fboxrule}{0.3pt}%
        \fbox{\includegraphics[clip=true,trim=0mm 3mm 0mm 0mm,width=\columnwidth]{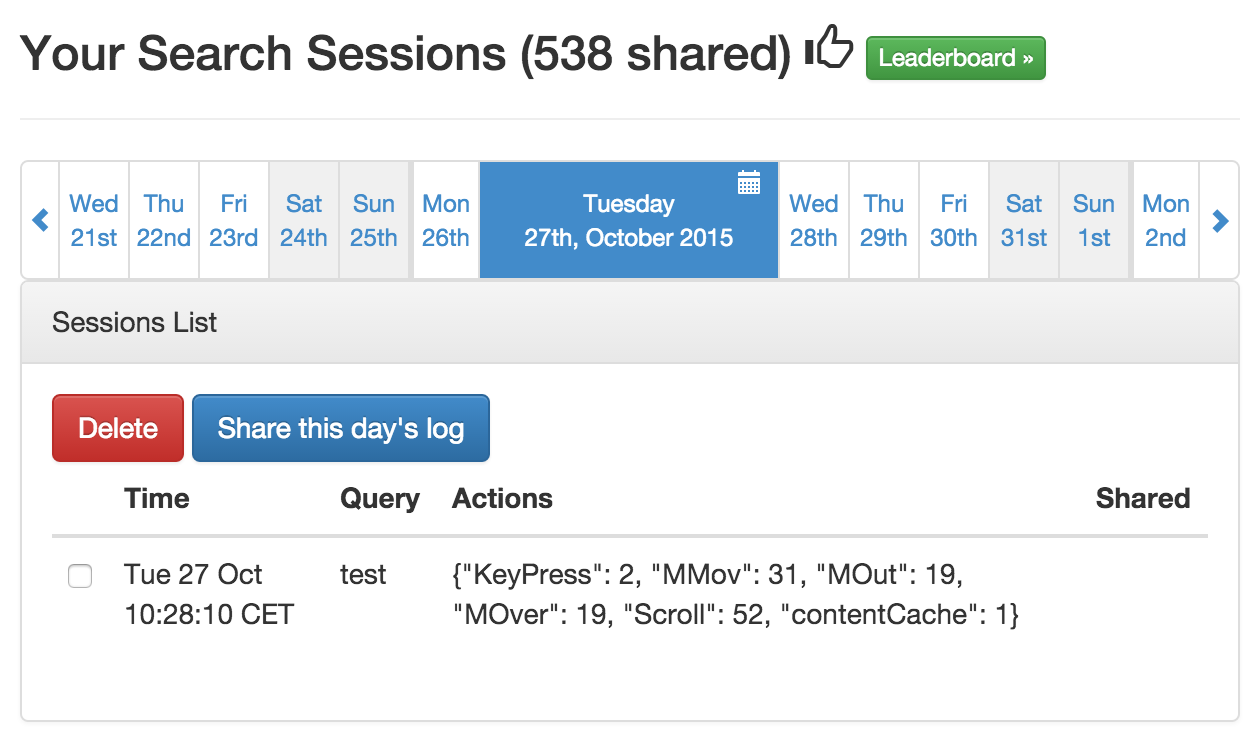}}%
    }
    \vspace*{1mm}
\caption{Log management interface for experiment participants.}
\label{fig:log_management}
\end{figure}

\begin{figure}
\centering
    \includegraphics[clip=true,trim=2.3mm 3.4mm 4mm 4.6mm,width=\columnwidth]{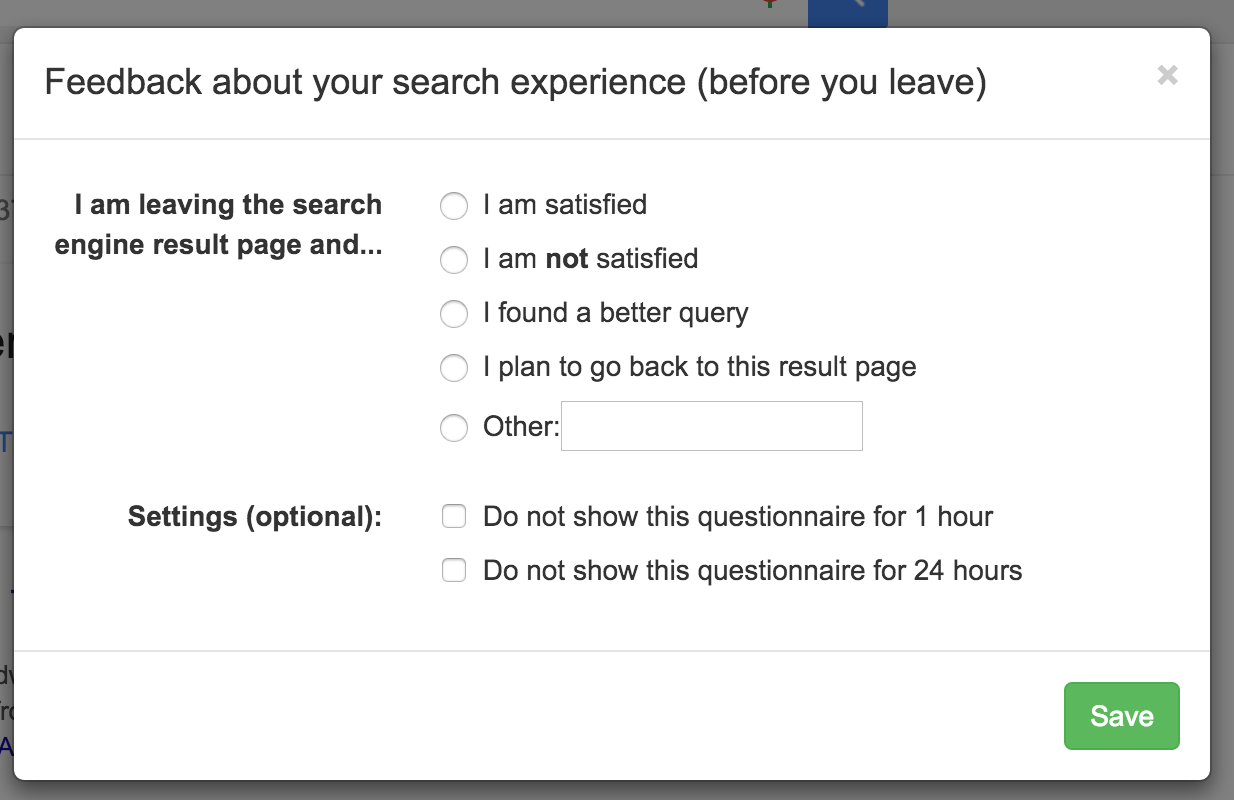}
    \vspace*{1mm}
\caption{Search satisfaction questionnaire.}
\label{fig:questionnaire}
\end{figure}

First of all, we set up a proxy search interface that intercepts user queries to a commercial
search engine and collects click and mouse movements data. The log collection code is
based on the \texttt{EMU.js} library by~\citet{Guo2008}.
The interface was used by a group of volunteers who agreed to donate their interaction data.
The design of the experiment was also reviewed by the University's Ethical Committee.
We only used the queries that were explicitly vetted by the owners as not privacy sensitive
using the log management interface we provide;
see Figure~\ref{fig:log_management}.\footnote{%
Our code, including modifications to \texttt{EMU.js} is available at \url{https://github.com/varepsilon/cas-eval}.}
We should also stress here, that unlike laboratory settings, the search experience was not changed:
the user received the same list of results and interacted with them in the same way
as if they were using the underlying search system in the normal manner.
Occasionally we showed a pop-up questionnaire asking users
to rate their search experience upon leaving the SERP\@;
see Figure~\ref{fig:questionnaire}.
To avoid showing it prematurely, we forced result clicks to open a new browser tab.
Through this questionnaire we collected explicit satisfaction labels that we later used as ground truth
to train and evaluate the CAS model.
Each user saw the pop-up questionnaire no more than ten times a day and only for $50\%$
of the sessions. The questionnaire was equipped with ``mute buttons'' that allowed the user to
disable the questions for one hour or $24$ hours.
We assumed that this questionnaire, if it was not shown overly frequently, would not
seriously affect the overall user experience. A similar setup was used in~\cite{Diriye2012}.

The dataset parameters are summarized in~Table~\ref{tab:proxydata}.
%We will also release the fully anonymized version of the dataset upon acceptance of the paper.
\begin{table}[h]
    \centering
    \caption{Data collected with the search proxy.}
    \begin{tabular}{lr}
        \toprule
        \# of participants & \num{12} \\
        \# of shared sessions (queries) & \num{2334} \\
        \# of shared sessions with satisfaction feedback & \num{243} \\
        %\# of shared sessions that were satisfactory & \num{148} \\
        %\# of shared sessions that were not satisfactory & \num{51} \\
        \bottomrule
    \end{tabular}
    \label{tab:proxydata}
\end{table}

% subsection proxydata (end)

\subsection{Crowdsourcing data collection}
\label{sub:cfdata}
As a second stage of our experiment
we asked crowdsourcing raters (``workers'') to assign (D) and (R) labels
(see Section~\ref{sec:metric})
by showing them SERP items or corresponding web documents and asking the following questions:
\begin{itemize}
    \item [(D)] Does the text above answer the question $Q$?
    %\item [(A)] Above is a summary extracted from a bigger document.
        %Do you think examining the full document will be useful to answer the question $Q$?
    \item [(R)] Does the document that you see after clicking the link contain an answer to the question $Q$?
\end{itemize}
For the first question we showed only the part of the SERP corresponding to
a single SERP item and no clickable links.
For the second one
we only showed a link and required the workers to click it.
Moreover, the above two tasks were run separately so the chances of
raters confusing the two tasks were quite low.
When comparing the most common (D) and (R)
labels assigned for each document, they show Pearson correlation values of $0.085$
and Spearman correlation values of only $0.094$, which proves that they are quite different.

Originally, a third question was also included to collect \emph{attractiveness}
labels (``(A)-ratings'') to be used instead of (R) relevance in \eqref{eq:alpha_logistic}.
It ran as follows: ``Above is a summary extracted from a bigger document.
Do you think examining the full document will be useful to answer the question $Q$?''
However, this proved to be a very subjective question,
and attractiveness labels collected this way were less useful as click predictor compared to relevance labels (R).
To be precise, the average (A)-rating for the clicked results was $0.82$,
while it was $0.84$ for non-clicked ($0.02$ standard deviation for both).
For the (R)-ratings the corresponding numbers were as follows: $2.29$ (standard deviation of $0.29$) for clicked
and $2.19$ (standard deviation of $0.31$) for non-clicked.
That proves that (R) serves better as a click predictor.
%\todo{Convert to a table and add small but nice
%\href{https://www.evernote.com/shard/s61/nl/1783680055/5f78029b-3a35-4b66-8b74-bfb92e951d6c/}{distribution plots}
%(if space permits)}.

From preliminary runs of the crowdsourcing experiment we learned that the crowd workers
rarely pay attention to the detailed instructions of a task, so we decided against using terms like
``query'' (we used \textit{question} instead) or ``snippet'' (we referred to it
as \textit{text} or \textit{summary}).
After several iterations of improving the task we also
decided to ask the raters to provide justifications for their answers.
We later used this as an additional signal to filter out spammers
(see Appendix~\ref{sec:filtering_spammers}),
but it can also be used to understand more about the complexity of individual
questions or the task as a whole~\cite{Aroyo2014}.
One application for the data collected in this way we already saw when we discussed
the satisfaction model in Section~\ref{sub:satisfaction_model}.
Another analysis that we ran was to identify potential good abandonments,
i.e., queries that may be answered directly on a SERP~\cite{Li2009}.
We found out that, even though the raters often disagree with themselves,\footnote{%
Approximately $30\%$ of the raters said that a query is both a potential good and bad abandonment when a slightly
different wording was used (or indicated that a potential bad abandonment query has an answer on a SERP).}
the queries that were marked as potential good abandonments
most often by the raters, were all labeled as such in an independent rating.

\begin{figure}[h]
\centering
    {%
        \setlength{\fboxsep}{0pt}%
        \setlength{\fboxrule}{0.3pt}%
        \fbox{\includegraphics[width=\columnwidth]{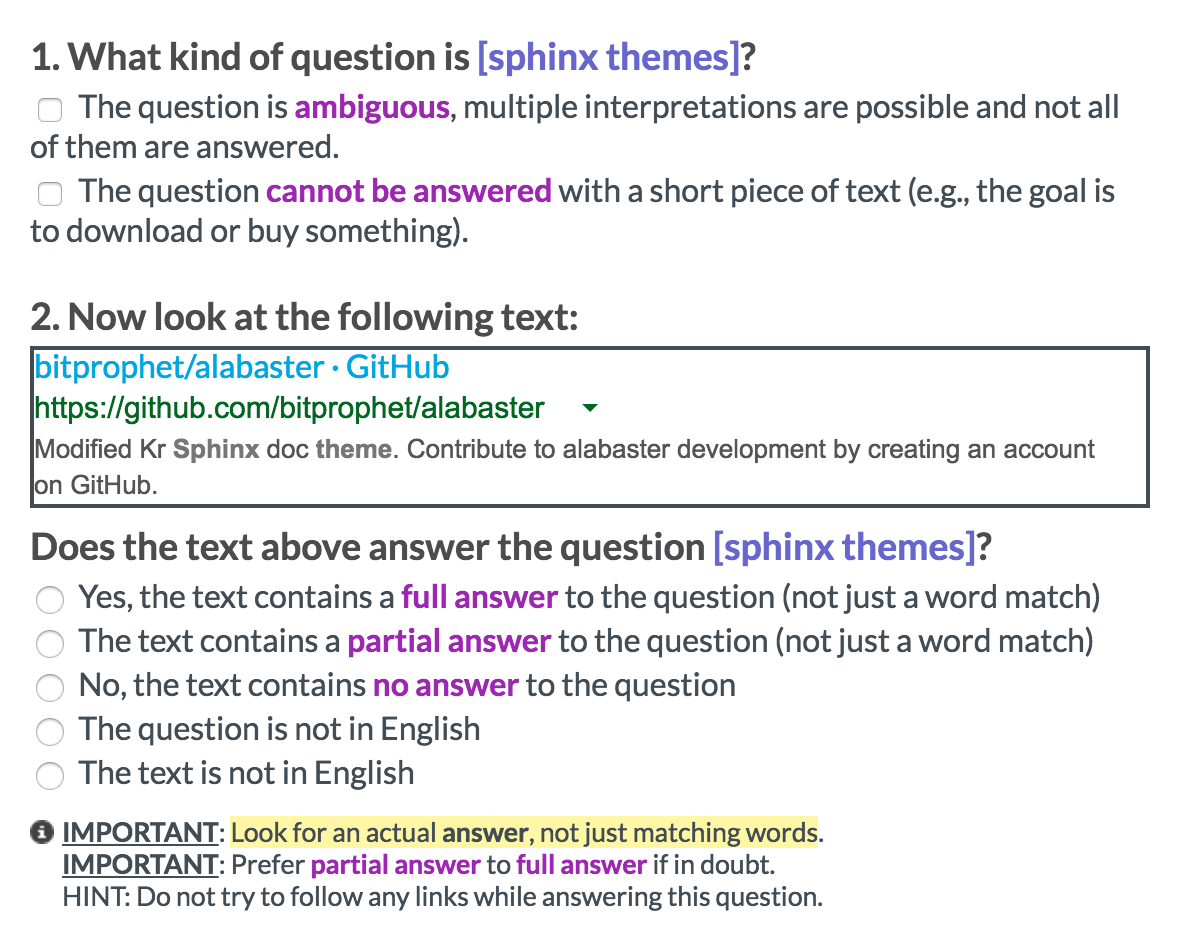}}%
    }
    \vspace*{1mm}
\caption{Crowdsourcing task for assigning direct relevance label (D) plus some additional questions.}
\label{fig:cf_task}
\end{figure}

An example of the task interface is shown in~Figure~\ref{fig:cf_task}.
We used the CrowdFlower platform, which is the only crowdsourcing platform we know of
that is available outside of the US\@. Workers were paid $\$0.02$ per task to keep the hourly
pay above $\$1$, well above the minimum wage of one of the author's home country
and a psychological threshold for the raters to treat it as a fair pay.\footnote{%
The workers were shown an optional survey at the end of the task
where they rated ``Pay'' from $3.2$ to $3.5$ (out of $5$).}

The key parameters of the dataset that we collected in this manner
are summarized in Table~\ref{tab:cfdata}.\footnote{%
The anonymized version of the dataset can be obtained
at \url{http://ilps.science.uva.nl/resources/cas-eval}.}
\begin{table}
    \centering
    \caption{Data collected via crowdsourcing.
    We sent for rating all the sessions with satisfaction feedback (Table~\ref{tab:proxydata}) apart from non-English queries.}
    \begin{tabular}{lrr}
        \toprule
                & \multicolumn{1}{c}{(D)}   & \multicolumn{1}{c}{(R)} \\
        \# of workers & \num{1822} & \num{951} \\
        \# of ratings & \num{23000} & \num{22056} \\
        \# of snippets/documents rated & \num{2180} & \num{2180} \\
        \bottomrule
    \end{tabular}
    \label{tab:cfdata}
\end{table}
After removing ratings coming from spammers (see Appendix~\ref{sec:filtering_spammers})
and sessions that are labeled as something other than ``I am satisfied''
or ``I am \textbf{not} satisfied'' (see Figure~\ref{fig:questionnaire})
we are left with $199$ query sessions.
Of those, $74\%$ were marked as satisfactory;
$12\%$ (24 items) of the SERPs are heterogeneous,
meaning that they have something other than ``ten blue links.''
For these $199$ queries we have $\num{1739}$ rated results. If an item does not have a rating,
we assume the lowest rating 0, although more advanced approaches exist~\cite{Aslam2007,Buttcher2007}.
% subsection cfdata (end)

\subsection{Baseline models/metrics}
\label{sub:baseline}
To compare the performance of our CAS model, we implemented the following baseline models:
\begin{itemize}
    \item the \textbf{UBM} click model by~\citet{Dupret2008} that was shown to be well correlated with user signals~\cite{Chuklin2013a};
    \item the \textbf{PBM} position-based model~\citep{Chuklin2015}, a robust model with fewer parameters than UBM\@;
    \item a \textbf{random} model that predicts click and satisfaction with fixed probabilities (learned from the data).
\end{itemize}

\noindent%
Apart from these, we also included the following metrics:
\begin{itemize}
    \item the \textbf{DCG} metric~\cite{Jarvelin2002} commonly used in IR evaluation~\cite{Jarvelin2002}; and
    \item the \textbf{uUBM} metric, the metric that showed the best results in~\cite{Chuklin2013a}.
        It is similar to the above UBM model, but parameters
        are trained on a different and much bigger dataset, namely a search log of Yandex.\footnote{%
\url{https://yandex.com}, the most used search engine in Russia.}
\end{itemize}

\noindent%
This way we include both non-model-based (DCG) and model-based metrics (the rest),
but also locally trained models (UBM, PBM) as well as the uUBM model trained on a different dataset.
% subsection baseline (end)

For testing we employ $5$-fold cross-validation that we restart $5$ times,
each time reshuffling the data, see Algorithm~\ref{alg:tqfold}.
Thus, we have $25$ experimental outcomes that we aggregate to assess significance of the results.
\begin{algorithm}
    \caption{$TQ$-fold cross-validation.}
    \label{alg:tqfold}
    \begin{algorithmic}[1]
        \Procedure{TQ-fold}{dataset $D$, $T$ repetitions, $Q$ folds}
        \State $N \gets \mathit{size}(D)$
        \For {$i \gets 1 \text{ to }T$}
            \State $D \gets \mathit{RandomShuffle}(D)$
            \For {$j \gets 1 \text{ to }Q$}
            \State $D_\text{test} \gets D\left[\frac{N}{Q}(i - 1) \ldots \frac{N}{Q}i \right]$
            \State $D_\text{train} \gets D \setminus D_\text{test}$
            \State train on $D_\text{train}$
            \State evaluate on $D_\text{test}$
            \EndFor
        \EndFor
        \EndProcedure
    \end{algorithmic}
\end{algorithm}

% section experiment (end)

%% file: cikm2016-evaluation-results.tex
% !TEX root = cikm2016-evaluation.tex

\section{Results}
\label{sec:results}
Below we report results on comparing the CAS model and corresponding
evaluation metric to other models and metrics, respectively.

\subsection{Evaluating the CAS model}
\label{sub:resultsmodel}
We evaluate the CAS model by comparing the log-likeli\-hood values for different events, viz.\ clicks and satisfaction.
We also analyse the contribution of different attention features introduced in~Table~\ref{tab:attentionfeatures}.

\begin{figure}
\begin{center}
    \includegraphics[clip=true,trim=8mm 0mm 12mm 0mm,width=\columnwidth]{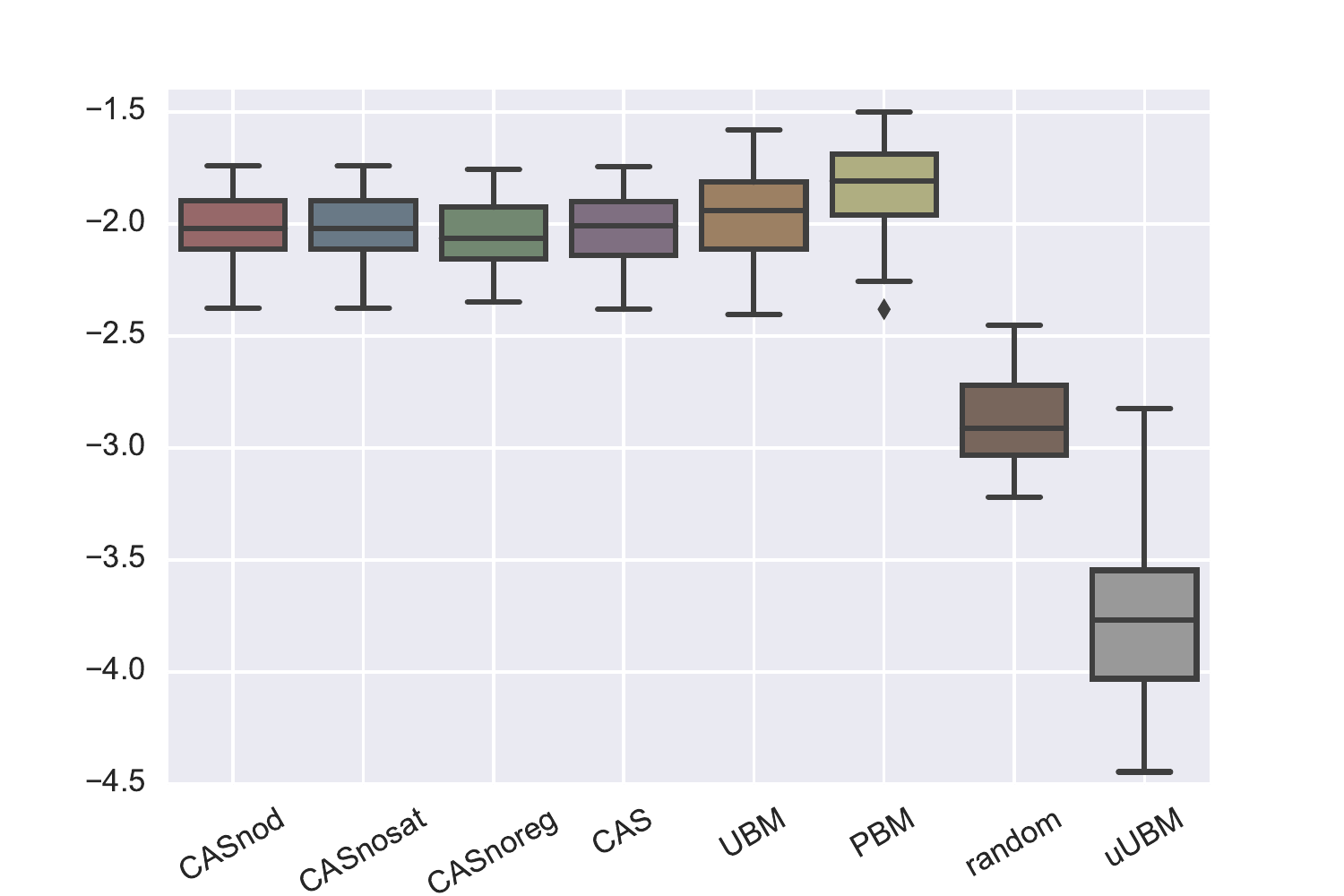}
\end{center}
\caption{Log-likelihood of the click data. Note that uUBM was trained on a totally different dataset.}
\label{fig:llclick}
\end{figure}

\paragraph{Likelihood of clicks}
First, we would like to know how the CAS model compares
to the baseline models in terms of log-likelihood.
Figure~\ref{fig:llclick} shows the likelihood of \emph{clicks}
for different models. On top of the CAS model described above,
we also included three modifications:
\begin{itemize}
    \item \textbf{CASnod} is a stripped-down version that does not use (D) labels;
    \item \textbf{CASnosat} is a version of the CAS model that does not include the satisfaction term~\eqref{eq:sat}
        while optimizing the model; and
    \item \textbf{CASnoreg} is a version of the CAS model that does not
        use regularization while training.\footnote{%
All other models were trained with $L_2$-regularization.}
\end{itemize}
As we can see from Figure~\ref{fig:llclick}, the difference between different variants of CAS is minimal in terms of click log-likelihood,
but we will see later that they are, in fact, different.
UBM and PBM show better log-likelihood values on average, with PBM being more robust.
There are two reasons for CAS to underperform here.
First, it is trained to optimize the full likelihood, which includes moused results and satisfaction,
not just the likelihood of clicks. As we will see later, CAS shows much better likelihood for satisfaction,
more than enough to make up for a slight loss in click likelihood.
Second, the class of models for examination and attractiveness probabilities
we have chosen (logistic regression) may not be
flexible enough compared to the arbitrary rating-to-probability mappings used by PBM and UBM\@.
While similar rating-to-probability mappings can be incorporated into CAS as well,
it makes the training process much harder and we leave it for future work.

\begin{figure}
\begin{center}
    \includegraphics[width=\columnwidth]{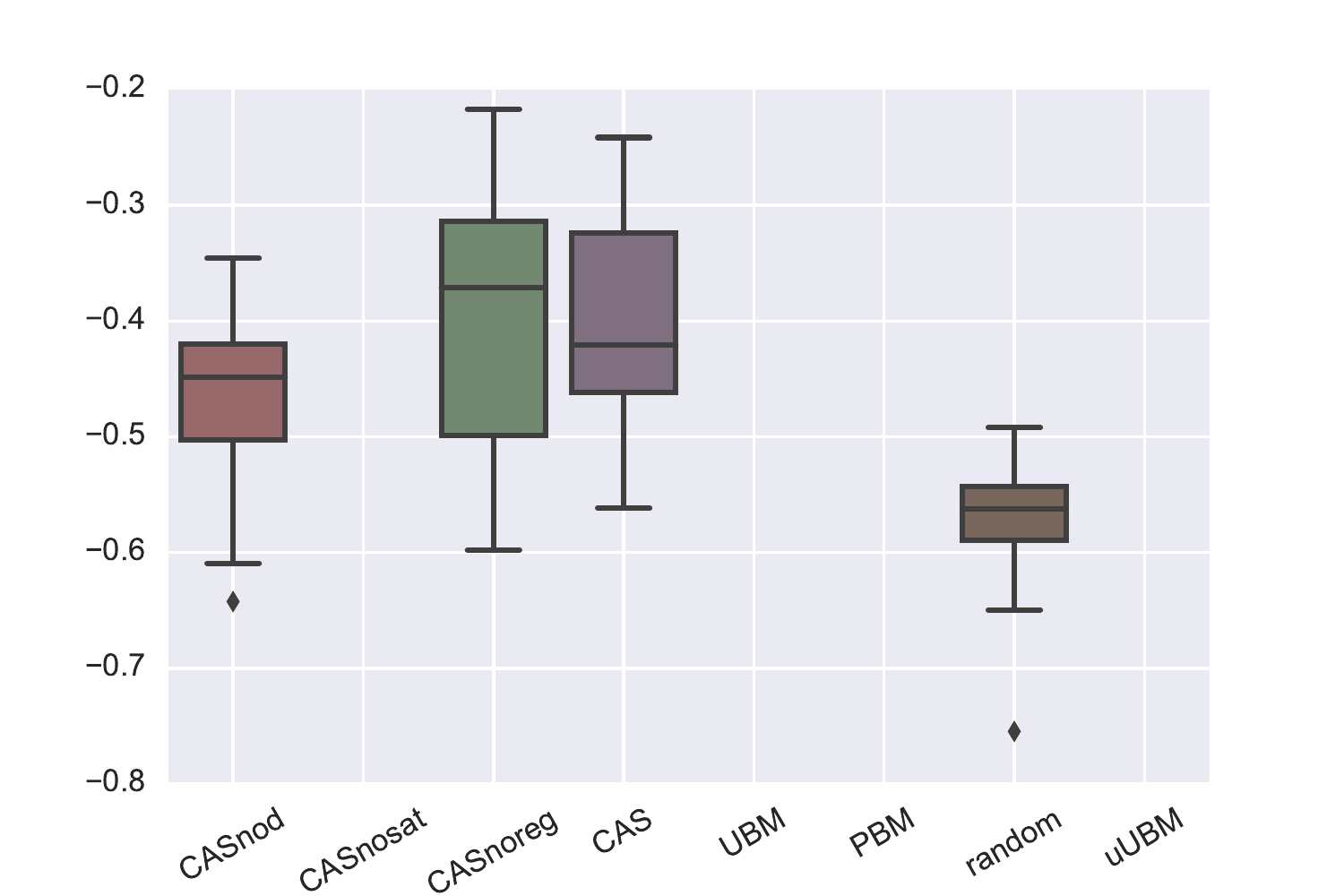}
\end{center}
\caption{Log-likelihood of the satisfaction prediction.
Some models here always have log-likelihood below $-0.8$, hence there are no boxes for them.}
\label{fig:llsat}
\end{figure}

\paragraph{Likelihood of satisfaction}
Next, we look into the log-likelihood of the satisfaction predicted by the various models;
see Figure~\ref{fig:llsat}. For the models
that do not have a notion of satisfaction (CASnosat, UBM, PBM, uUBM),
we used the sigmoid transformation of the utility function, which, in turn, was computed as
the expected sum of relevance of clicked results (see~\cite{Chuklin2013a}).
However, all such models were inferior to the random baseline; this finding supports the idea of
collecting satisfaction feedback directly from the user instead of relying on an ad-hoc interpretation
of utility that may be quite different from the user's perception of satisfaction.

By comparing the results for CAS vs.~CASnoreg in Figure~\ref{fig:llsat} we also see that regularization leads to a more stable satisfaction prediction likelihood, which is, however, lower on average.
If we have a large sample of data that is representative of the user population, regularization may as well be omitted.
By comparing the performance of CAS vs.~CASnod we can also see that the lack of (D) ratings clearly hurts the model's performance as it now cannot explain some of the
utility directly gained from the SERP.

\paragraph{Analyzing the attention features}
Finally we look at the features used by the attention model (Table~\ref{tab:attentionfeatures}).
If we exclude some of these features we obtain the following simplified versions of the CAS model:
\begin{itemize}
    \item \textbf{CASrank} is the model that only uses the rank to predict attention;
        this makes the attention model very similar to PBM and the existence of
        the satisfaction component~\eqref{eq:sat} is what makes the biggest difference;
    \item \textbf{CASnogeom} is the model that only uses the rank and SERP item type information and does
        not use geometry; and
    \item \textbf{CASnoclass} is the model that does not use the CSS class features (SERP item type).
\end{itemize}
We compare these models to the vanilla CAS and CASnod models in terms of log-likelihood of click
and satisfaction prediction as we did above for the baseline models.

\begin{figure}
    \centering
    \subfloat[Clicks.]{\includegraphics[width=0.23\textwidth]{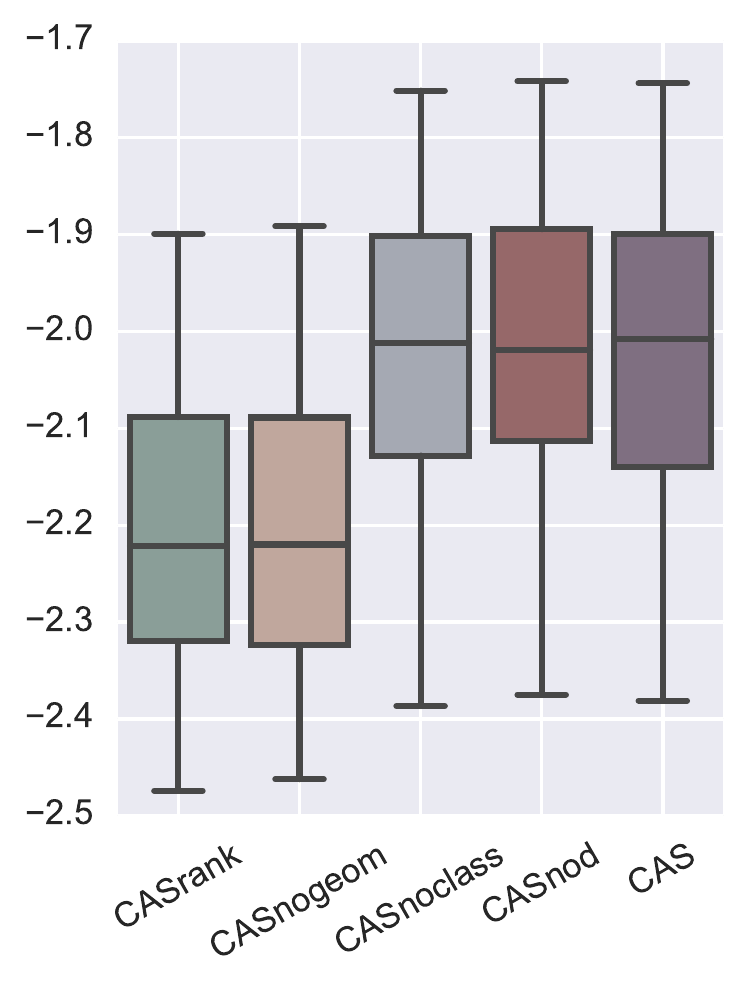}\label{fig:llclickatt}}
    \hfill
    \subfloat[Satisfaction.]{\includegraphics[width=0.23\textwidth]{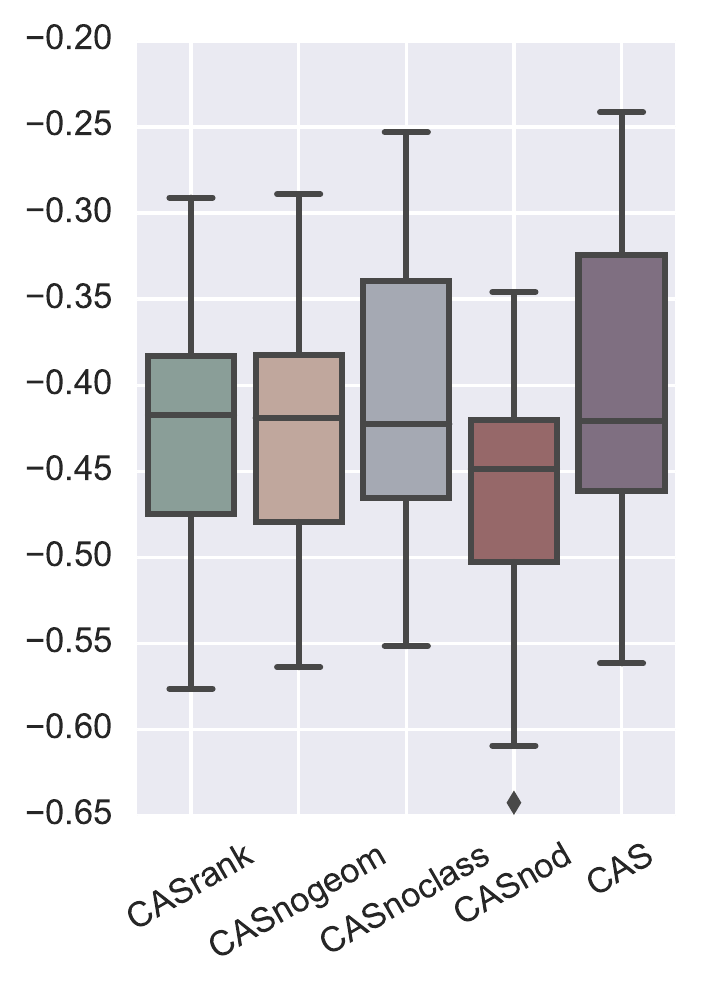}\label{fig:llsatatt}}
    \vspace*{2mm}
    \caption{Feature ablation for the attention model: Log-likelihood of the click prediction (a) and
    the satisfaction prediction (b) for vanilla CAS as well as stripped-down versions of it.}
\end{figure}

The results are shown in Figures~\ref{fig:llclickatt}~and~\ref{fig:llsatatt}.
What we can see from these plots is that excluding (D) labels (CASnod) almost
does not affect click prediction accuracy,
but it does substantially hurt the satisfaction prediction.
This is expected as these labels are only used in the
satisfaction formula~\eqref{eq:sat}. On the other hand, removing geometry features (CASnogeom, CASrank)
hurts click prediction the most, while having a less prominent impact on satisfaction prediction.
Finally, removing CSS class features (CASnoclass) has a small effect on both click and satisfaction prediction,
but much smaller than removing geometry affects click prediction or removing (D) labels affects satisfaction prediction.

\medskip\noindent%
In this section we showed that the CAS model predicts clicks slightly worse than the baseline models,
albeit at roughly the same level. When it comes to predicting satisfaction events, the baseline models show
much lower log-likelihood values, the only comparable performance is shown by the random model,
but it still performs worse than CAS\@.
In terms of incorporating satisfaction into our models, we demonstrated that it is
necessary to do so in order to beat the random baseline on the log-likelihood of satisfaction
(CASnosat is always worse than the baseline) and the (D)-labels play an essential role for model accuracy:
CASnod shows lower log-likelihood than the CAS\@. This answers our first research question~\ref{rq1}.

% subsection resultsmodel (end)

\subsection{Evaluating the CAS metric}
\label{sub:resultsmetric}

Now we evaluate the metric derived from the CAS model and described in~Section~\ref{sec:metric}.
To do this we compute correlations with baseline metrics and with user-reported satisfaction.

\paragraph{Correlation between metrics}
Table~\ref{tab:pearson} shows the average Pearson correlation between utilities produced by different metrics averaged
across folds and repetitions of cross-validation.
\begin{table}
    \centering
    \caption{Correlation between metrics measured by average Pearson's correlation coefficient.}
    \label{tab:pearson}
    \small
    \begin{tabular}{@{}l@{~}c@{~}c@{~}c@{~~}c@{~~}c@{~~}c@{~~}c@{}}
    \toprule
    {} &  \mbox{}\hspace*{-3mm}CASnosat &  CASnoreg &   CAS &   UBM &   PBM &   DCG &  uUBM \\
    \midrule
    CASnod   &     0.593 &     0.564 & 0.633 & 0.470 & 0.487 & 0.546 & 0.441 \\
    CASnosat &          &     0.664 & 0.715 & 0.707 & 0.668 & 0.735 & 0.684 \\
    CASnoreg &          &          & 0.974 & 0.363 & 0.379 & 0.417 & 0.341 \\
    CAS      &          &          &      & 0.377 & 0.394 & 0.440 & 0.360 \\
    \midrule
    UBM      &          &          &      &      & 0.814 & 0.972 & 0.882 \\
    PBM      &          &          &      &      &      & 0.906 & 0.965 \\
    DCG      &          &          &      &      &      &      & 0.943 \\
    \bottomrule
    \end{tabular}
\end{table}
As we can see, metrics from the CAS family are less correlated with the baseline metrics than they are with each other.
The highest level of correlation with the baseline metrics among the CAS metrics is achieved by CASnosat,
the metric that does not explicitly include satisfaction in the user model.
This is expected as its model is close to PBM\@.
Another observation from Table~\ref{tab:pearson} is that CASnod is also quite different
from the baseline metrics, but not as much as CASnoreg and CAS, which, again, shows that including (D)
relevance labels (direct snippet relevance) makes the metric quite different.

\begin{figure}
\begin{center}
    \includegraphics[clip=true,trim=8mm 0mm 12mm 0mm,width=\columnwidth]{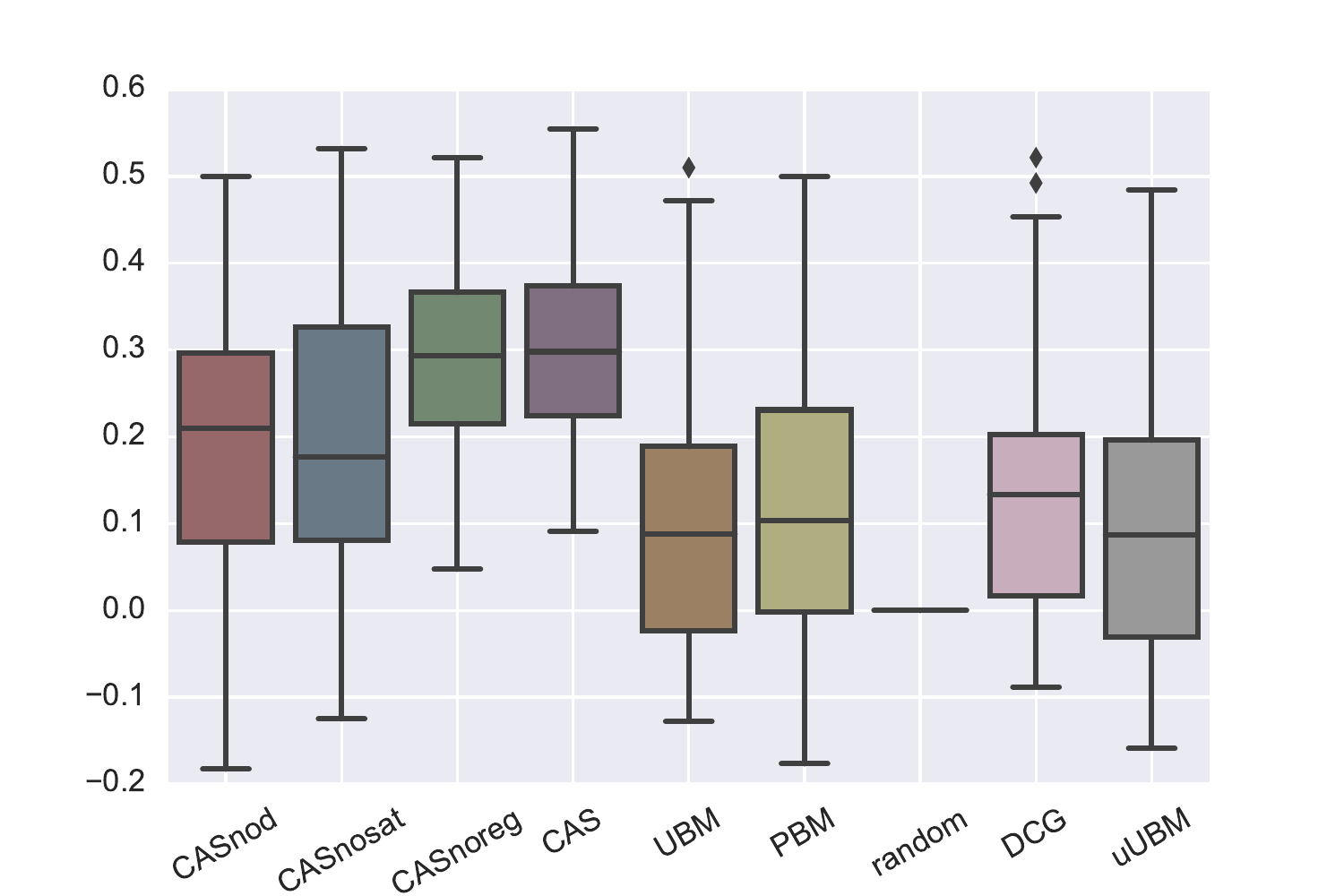}
\end{center}
\caption{Pearson correlation coefficient between different model-based metrics and the user-reported satisfaction.}
\label{fig:satpearson}
\end{figure}

\paragraph{Correlation with user-reported satisfaction}
Figure~\ref{fig:satpearson} shows the Pearson correlation between the utility induced by one of the models
and the satisfaction reported by the user (zero or one).
As we can see from the plot, the metric induced by the CAS model shows the best Pearson correlation values,
despite the fact that it was trained to maximize the full likelihood of the data,
not just to predict satisfaction.
Correlation is always above zero for metrics based on CAS and CASnoreg, but for the metrics based on CASnod and CASnosat the correlation can be negative,
which, again, reinforces the importance of the (D) labels and the explicit satisfaction component in the model.
While comparing CAS to the baseline models,
we observed that the correlation values for the CAS-based metrics are at least $0.14$ higher on average.

To prove that the CAS model is especially useful in case of heterogeneous SERPs we performed the
following experiment. We made a stratified random split of the dataset into training and testing,
where the test set contains $1/24$ of the data and exactly one heterogeneous SERP
(as we mentioned in~Section~\ref{sub:cfdata}, our dataset contains $24$ such SERPs).
We then computed utility of this one SERP using the metric trained on the train set and
compared it to the satisfaction label for the corresponding session.
We then repeated this process $20$ times
and computed the Pearson correlation of the utilities and satisfaction labels.
Results are reported in~Table~\ref{tab:heterogeneousserps}.
We see that metrics of the CAS family show much higher correlation with
the user-reported satisfaction then other metrics.
\begin{table}[h]
    \small
    \centering
    \caption{Pearson correlation between utility of heterogeneous SERP and user-reported satisfaction.}
    \begin{tabular}{@{~}c@{~}c@{~}c@{~}c@{~}c@{~}c@{~}}
        \toprule
        CAS & UBM & PBM & random & DCG & uUBM \\
        \midrule
        0.60 & 0.38 & -0.05 & -0.39 & 0.24 & -0.08 \\
        \bottomrule
        \toprule
        CASrank & CASnogeom & CASclass & CASnod & CASnosat & CASnoreg \\
        \midrule
        0.15 & -0.04 & 0.27 & -0.04 & 0.48 & 0.67 \\
        \bottomrule
    \end{tabular}
    \label{tab:heterogeneousserps}
\end{table}

%\todo{Describe experiments with increasing the weight of satisfaction likelihood and NOT getting better results.}

\begin{figure}
\begin{center}
    \includegraphics[width=\columnwidth]{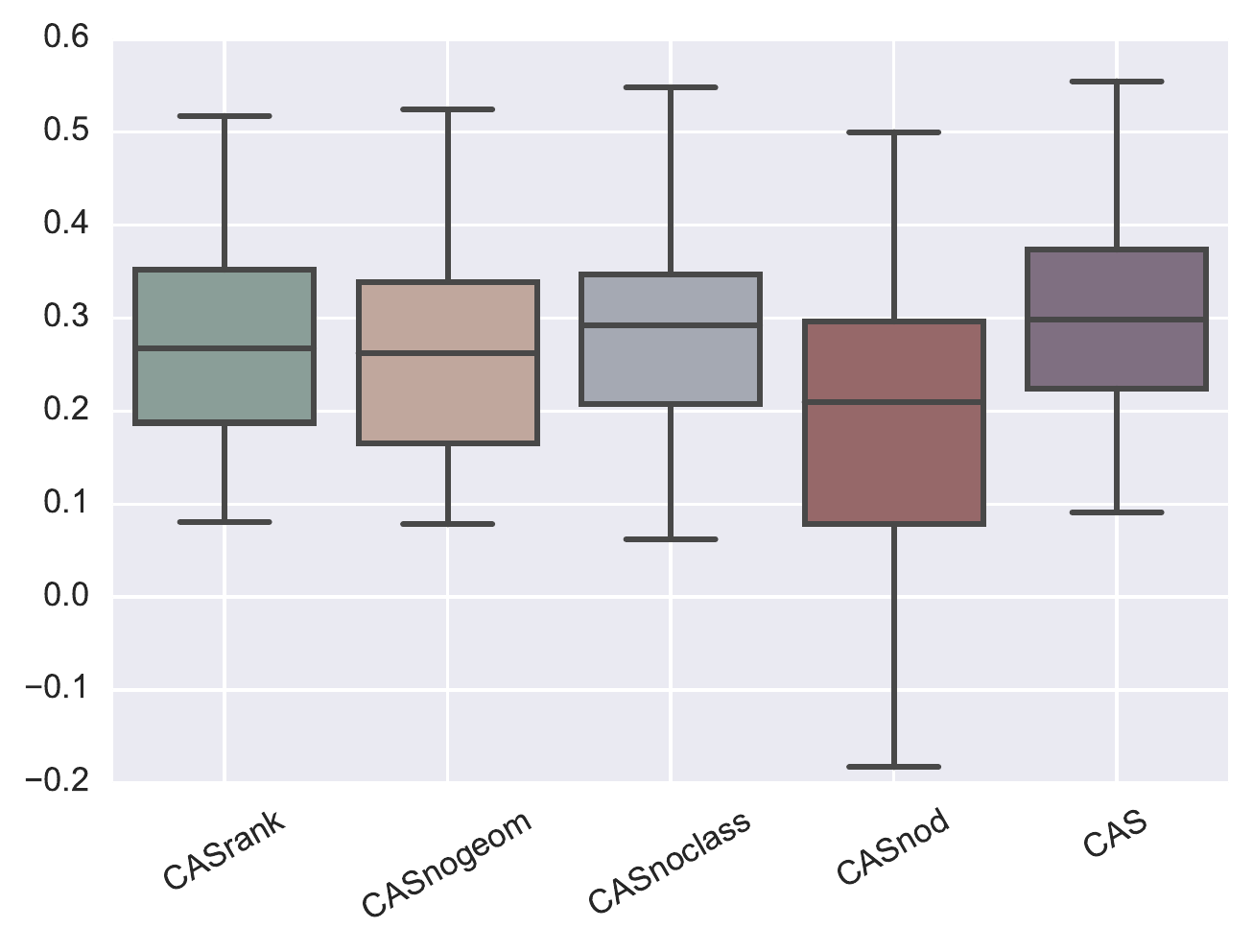}
\end{center}
\caption{Feature ablation for the attention model: Pearson correlation coefficient between different variants of the CAS metric and users' satisfaction.}
\label{fig:satpearsonattention}
\end{figure}

\paragraph{Analyzing the attention features}
Similar to our analysis in Section~\ref{sub:resultsmodel} we perform an ablation study, this time
to compare vanilla CAS to CASrank, CASnogeom, CASnoclass and CASnod
in terms of how well the metric induced by them is correlated with user-reported satisfaction.
The results are shown in Figure~\ref{fig:satpearsonattention}.

As can be see from the plot, removing the class features reduces correlation only a little (CAS vs.~CASnoclass).
We hypothesize that the reason for this is that in our dataset only $12\%$ of the SERPs have non-trivial SERP
items. Removing geometry features (CASnogeom) or both geometry and class features (CASrank)
already makes the metric perform worse, which supports the fact that modern SERPs require a
non-trivial attention model. Finally, the worst performing metric is CASnod which does not
use the (D)-labels. The performance drop is much higher than for the models discussed above,
which shows that attention features are important for satisfaction prediction,
but having (D)-labels brings more to the table.
This is consistent with the analysis of the results reported in Figure~\ref{fig:llsatatt}.

\medskip\noindent%
In this section we showed that the metric based on the CAS model that we propose differs substantially
from the baseline metrics, but
less so if the model does not include (D) labels or disregards the satisfaction term altogether.
More importantly, the CAS metric is not just different from the baseline metrics, it also shows better correlation
with the satisfaction reported by users. So, indeed, incorporating satisfaction yields a new and interesting metric,
which answers~\ref{rq2}.
% subsection resultsmetric (end)

% section results (end)

%% file: cikm2016-evaluation-discussion.tex
% !TEX root = cikm2016-evaluation.tex

\section{Discussion}
\label{sec:discussion}
First of all, we would like to acknowledge some limitations of the paper.
Our dataset is small compared to the typical datasets used for training click models~\cite{Chuklin2015}
and may be somewhat biased in terms of query distribution since most of the users
whose data was used have a Computer Science background.
It would be preferable to collect such data at a bigger scale. One direction for future
work would be to train the CAS model on heterogeneous data, where potentially a bigger
dataset with clicks and mousing is supplemented by a smaller one with satisfaction labels.

Feature engineering for the attention model is also not comprehensive and was not a goal
of the current paper. One may add more saliency features to detect the users' attention
or even train separate skimming and reading models~\cite{Liu2014}.

Another challenging part in our setup is the use of crowd workers. It would be interesting
to run a study with trained raters and learn how to extrapolate it to the crowd,
by adjusting the instructions and filtering the spammers in a more automated fashion than we have used~\cite{Kazai2016}.
There is also a noticeable difference between raters and the users. For example,
\citet{Liu2015} claim that the raters pay more attention to the effort required to complete a task,
while the users care more about utility. Also, the ratings assigned by the owners of the query
are different from the ones assigned by other people~\cite{Chouldechova2013}.

Mobile search evaluation~\cite{Guo2013,Huang2012b}
is another facet of future work. As we mentioned before,
navigating away from a SERP is more expensive there, so the users tend to gain
utility directly from the SERP and the search engines add more ways to help this.
It would be interesting to see how we can leverage additional attention signals
to adapt the CAS model for mobile settings.
% section discussion (end)

%% file: cikm2016-evaluation-conclusion.tex
% !TEX root = cikm2016-evaluation.tex

\section{Conclusion}
\label{sec:conclusion}
In this paper we have presented a model of user behavior that combines clicks, attractiveness and satisfaction in a joint model,
which we call the CAS model. We have also proposed a method to estimate the parameters of the model and
have shown how a Cranfield-style offline evaluation metric can be built on top of this model.
We have also described the crowdsourcing setup to collect labels for individual documents.

We have demonstrated that \emph{the model} conceived in this way can be used as a robust
predictor of user satisfaction without sacrificing its ability to predict clicks.
We have also shown that decoupling satisfaction from attention and clicks leads to
inferior satisfaction prediction without gain in predicting clicks.

In addition, we have presented \emph{a metric} that can be used for offline search system evaluation,
an important component of ranking development. The CAS metric with parameters
trained from user data consistently shows correlation with satisfaction,
unlike traditional metrics. Moreover, the metric is quite different, suggesting
that including it into one's evaluation suite may lead to a different view on which
version of the ranking system is better.

While the current study has its limitations, we view it as a motivation to
move away from the ten blue links approach and adopt an evaluation metric
that uses rich features and relevance signals beyond traditional document relevance.
We also call for releasing a dataset that would allow
for a more comprehensive evaluation than currently provided by TREC-style evaluation setups\@.
% section conclusion (end)

%% file: cikm2016-evaluation-appendix.tex
% !TEX root = cikm2016-evaluation.tex

\section{Filtering Spammers}
\label{sec:filtering_spammers}
%\small
To identify spammers, we used the free-text fields where the raters were asked to copy text
from the snippet or full document to support their relevance ratings.
If the text was not copied from the snippet
(in case of direct snippet relevance) or contained gibberish words, we added this
worker to the list of suspicious workers.
After each batch of tasks sent for ratings was finished, we manually reviewed top lowest scoring
workers according to those metrics and banned them from the future tasks.\footnote{%
Manual examination of the worker's ratings before banning is enforced by the crowdsourcing platform that we used.
} We also ignored workers with fewer than three ratings following~\cite{Aroyo2014}.
In total, we ignored ratings coming from $698$ workers out of $2185$, which corresponds to $27\%$
of direct snippet relevance ratings (D) and $29\%$ of relevance ratings (R).

To measure worker disagreement we reported average Cohen's kappa~\cite{Cohen1960}
as well as Krippendorf's alpha~\cite{Krippendorff1970}.
The numbers are reported in~Table~\ref{tab:spammers}.
\begin{table}
    \centering
    \caption{Filtered out workers and agreement scores for remainging workers.}
    \begin{tabular}{l@{~}c@{~~}c@{~~}c@{~~}c}
        \toprule
           & \% of workers & \% of ratings & Cohen's & Krippendorf's \\
        label   &         removed &       removed &    kappa & alpha \\
        \midrule
        (D)  & 32\% & 27\% & 0.339 & 0.144 \\
        (R)  & 41\% & 29\% & 0.348 & 0.117 \\
        \bottomrule
    \end{tabular}
    \label{tab:spammers}
\end{table}
As we can see, the agreement numbers are rather low, which shows that there is
still a big variation of opinions. Fortunately, our model is able to accommodate
this by taking the histogram of ratings and not just a single number
coming from averaging or majority vote,
see~equations~\eqref{eq:u_d},~\eqref{eq:u_r}~and~\eqref{eq:alpha_logistic}.

We also experimented with worker-worker and worker-task disagreement scores~\cite{Aroyo2014}.
We remove workers that disagree with too many other workers on either global
or per-item level. We explored different thresholds on disagreement scores and managed
to improve overall agreement measured by Cohen's kappa and Krippendorf's alpha
(which was expected) but it did not improve the results (Section~\ref{sec:results}).
We suspect that the reason is that there are always enough careless workers
that consistently give wrong answers and show good agreement with each other,
and the worker disagree\-ment-based method is not able to catch them.
Moreover, some disagreement is natural in such a subjective task; reducing it
does not necessarily improve quality.

Finding a good balance between data quantity and data quality
is a topic for a different discussion and is outside the scope
of this paper. Changing
the settings for spammer filtering to very aggressive
(removing all workers that made at least one mistake thus filtering
out over $75\%$ of the ratings)
and to very permitting (no spammer filtering) both give rise to
models with inferior performance both in terms of likelihood of clicks/satisfaction
and the correlation of utility and user-reported satisfaction.
%\todo{Show the numbers in the results section, refer from here}.
% section filtering_spammers (end)